# Triplet Superconductors from the Viewpoint of Basic Elements for Quantum Computers

Armen M. Gulian and Kent S. Wood

*Abstract*— We discuss possibilities of utilizing superconductors with Cooper condensates in triplet pairing states (where the spin of condensate pairs is S=1) for practical realization of quantum computers. Superconductors with triplet pairing condensates have features that are unique and cannot be found in the usual (singlet pairing, S=0) superconductors. The symmetry of the order parameter in some triplet superconductors (e.g., ruthenates) corresponds to doubly-degenerate chiral states. These states can serve as qubit base states for quantum computing.

*Index Terms*—Qubit, ruthenate, chiral states, quantum computing.

## I. INTRODUCTION

QUBITS are the heart of Quantum Computing (QC). As with the conventional bits that underlie classic computers by allowing two discrete states, 0 or 1, qubits in the simplest case can be in two discrete quantum states, conventionally $|0>$ and $|1>$. Between measurements they are represented as a superposition of both states: $|\Sigma> = \alpha|0> + \beta|1>$. These classically undetectable complex values of $\alpha$ and $\beta$ are manipulated during QC operations, eventually providing computational results by means of interaction with other qubits, via gates. The search for suitable physical qubits is ongoing: there is not yet a clear front-runner technology.

In considering a new candidate technology one first describes how it will realize the qubit and what stability and coherence properties it will have. Next one must present candidate techniques for how primitive operations will be carried out. This process admits a physical object to candidacy as a qubit. Presentation of the qubit concept and the approach for certain operational primitives are necessary and sufficient conditions for admitting a new technology to candidacy for QC. Our intent in this paper is to introduce triplet-pairing states in superconducting ruthenates as meriting consideration for quantum computing.

## II. QUBITS

### A. Quantum Properties of Small Ruthenate Samples

Triplet superconductors, with the Cooper pair spin $S=1$, corresponding to *p*-wave pairing as in superfluid $^3He$. Since $Sr_2RuO_4$ [1]-[3] is regarded as a "textbook example" of spin-triplet pairing, we refer primarily to that material during this discussion. Other relevant candidates could be found among "heavy fermion" superconductors such as $UBe_{13}$ [4], as well as quasi-one dimensional organic superconductors, such as Bechgaard's salts based on the tetramethyltetraselenafulvalene molecule, specifically $(TMTSF)_2PF_6$ [5].

Competition between ferromagnetic and antiferromagnetic fluctuations in $Sr_2RuO_4$ [6] favors triplet pairing with total spin $S=1$ [7]. $Sr_2RuO_4$ reveals a pronounced two-dimensional structure of the electron liquid (conductivity along the crystallographic *c*-axis is more than *800* times less than conductivity within the *(a,b)*-plane [1]). To describe the order parameter in ruthenates, one should analyze triplet superconductivity in two dimensions on a square lattice (in view of its tetragonal crystalline symmetry). The order parameter in triplet superconductors can be represented as $\hat{\Delta} = i\mathbf{d}(\mathbf{k}) \cdot \boldsymbol{\sigma} \sigma_y$, where the vector $\mathbf{d}$ has components $d_\alpha = A_{\alpha i}\hat{k}_i$ and $A_{\alpha i}$ is a $3 \times 3$-matrix of complex numbers (the indices $\alpha$ and $i$ correspond to the directions in spin and orbital (coordinate) space correspondingly, see, e.g., [8]-[10]). Experimental findings in ruthenates (in particular, the muon experiments [11]) point out that its superconducting state corresponds to the broken time-reversal symmetry. A corresponding analogy is $^3He-A$, the so-called A-phase of $^3He$ [12], where the energy gap has a nodal structure ($\Delta(\mathbf{k})=0$ at opposite poles on the Fermi-sphere). These poles specify a direction of anisotropy. The spin of Cooper pairs has only two non-zero projections: $S_z=1$ and -1. In ruthenates, because of finite spin-orbit coupling, neither spin nor orbital moments are good quantum variables. Possible pairing states, including splitting due to the spin-orbit coupling, are usually classified by irreducible representations $\Gamma_1^- - \Gamma_5^-$ [7]. The $\Gamma_5^-$ representation implies a symmetry for the wavefunction $\mathbf{d} = \hat{\mathbf{z}}\Delta(k_x \pm ik_y)/k_F$ (with total angular momentum *J=1*,

$J_z=\pm 1$ where $\hat{x}, \hat{y}, \hat{z}$ denote the directions in the spin-space) with $\hat{z}$ orthogonal to the conductivity plane. In the two-dimensional geometry (ruthenates) this state is nodeless, but it is still analogous to the A-phase of $^3$He, since two polar nodes disappear at infinite stretching of the Fermi-surface. Broken time-reversal symmetry means that the ground state of ruthenates should be doubly degenerate. Different *chiral states* (*N=1* and *N=-1*) are related to this degeneracy (see Fig. 1).

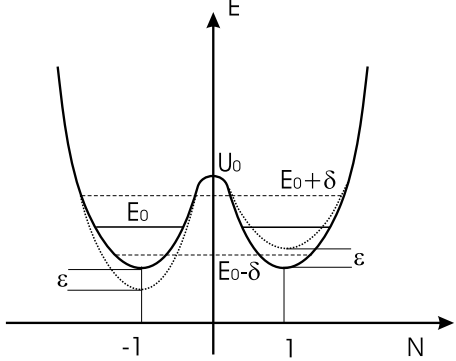

Fig. 1. Degeneracy in ruthenate samples related to chirality N. When the quantum wave-function is collapsed, the qubit is either in state |-1> or |1> of the phase space with the energy $E_0$ in each of them. In a *coherent* symmetric state [$2^{-1/2}$(|-1> + |1>)] due to tunneling the energy is $E_0-\delta$, and in anti-symmetric state [$2^{-1/2}$(|-1> - |1>)] — $E_0+\delta$. Transition between these two states should be possible with the emission or absorption of a photon. External fields can lift off the degeneracy (dotted line).

Chirality is characterized by a topological number $N = \frac{1}{4\pi} \iint dk_x dk_y \hat{\mathbf{m}} \cdot \left( \frac{\partial \hat{\mathbf{m}}}{\partial k_x} \times \frac{\partial \hat{\mathbf{m}}}{\partial k_y} \right)$, where the unit vector $\hat{\mathbf{m}} = \frac{\mathbf{m}}{|\mathbf{m}|}$, $\mathbf{m} = (\text{Re}\, d_z, \text{Im}\, d_z, \varepsilon_k)$, $\varepsilon_k = \frac{(k_x^2 + k_y^2)}{2m} - \mu$, and $\mu$ is the chemical potential [13]. For the state $\Gamma_5^-: N = \pm 1$, and in absence of external fields, $E(N=1)=E(N=-1)$. The duality of *N* implies the possibility of multiple domains. As usual with broken symmetries, in each domain a certain value of N is established at cooling down of the superconductor because of fluctuating initial parameters, and it may be different from cooling to cooling, as in $^3$He [14]. Large samples will have multi-domain structure. Small enough samples should constitute a single domain. Regarding superconductors as quantum objects, one can expect having quantum superposition of states *|N=1>* and *|N=-1>* in the same domain: $|\Sigma>=\alpha|1> + \beta|-1>$, so that $|\alpha|^2+|\beta|^2=1$. For equal energy states this superposition is a consequence of quantum tunneling. To understand better the consequences let us analyze this idea in some detail.

The 'spin-boson' Hamiltonian which corresponds to the situation described in Fig. 1 can be written in a form:

$$H_{SB} = E_0 \hat{I} - \delta \hat{\sigma}_x + \varepsilon \hat{\sigma}_z + H_{\text{int}} \qquad (1)$$

where $\delta$ corresponds to tunneling, $\varepsilon$ is a 'tuning' parameter, $H_{\text{int}}$ contains an interaction of tunneling system with the environment. Let us ignore for the moment the interaction with environment and put also $\varepsilon=0$. Then

$$\hat{H} = \begin{pmatrix} E_0 & -\delta \\ -\delta & E_0 \end{pmatrix} = E_0 \hat{I} - \delta \hat{\sigma}_x \qquad (2)$$

The solution of Schrödinger's equation has then a form (see, e.g., [15], we introduce $E_1=E_0-\delta$, $E_2=E_0+\delta$, and put $\hbar=1$):

$$|1> = (a/2)\exp[-(iE_1 t)] + (b/2)\exp[-i(E_2)t]$$
$$|-1> = (a/2)\exp[-(iE_1 t)] - (b/2)\exp[-(iE_2)t] \qquad (3)$$

If *b=0* system is at a minimal energy state $E=E_1$ (symmetric configuration) and if *a=0* – at a higher energy state $E=E_2$ (anti-symmetric configuration). These two states are the eigenstates of the Hamiltonian (2). If system is in one of these states, it can stay there indefinitely long in absence of perturbations.

The situation is less static when the system initially is in a 'collapsed' state, either in *|1>* or in *|-1>*. For that case, we should substitute *t=0* in Eqs. (3), getting *a/2+b/2=1, a/2-b/2=0*, so that *a=b=1*. This immediately yields a well-known result: starting at t>0 the system will coherently oscillate between states *|1>* and *|-1>*, so that the corresponding probability difference $P(t) = P_{|1>} - P_{|-1>} = Cos(2\delta t)$ (quantum beating with the frequency $E_2-E_1$). As soon as $H_{\text{int}}\neq 0$ the frequency $\delta$ will be renormalized, and the damping will take place [16]. The role of damping can be negligible in some cases, and crucial in other cases, so that overdamping will preclude any oscillation.

*B. Experimental Implications*

To be able to make a certain prediction a proper model for tunneling should be chosen. Another way to find an answer is to look into the closely related system $^3$He-A, where coherent oscillations between chiral states were detected experimentally [14], [17]. There is an analogy with instantons in describing coherent oscillations of the orbital moment between states **L** and –**L** [12], [18]. It is well known that in this experiment the state must be prepared by switching on the magnetic field in specific directions. Oscillations of the orbital momentum begin as soon as the field is off. The height of the potential barrier (Fig. 1) has a direct influence onto the frequency: the period of oscillations rapidly becomes large when T drops far below $T_c$. The same should be expected in ruthenates, but one needs a special investigation to make quantitative predictions. Regarding qualitative consequences, recently, a spontaneous Hall effect was predicted in ruthenates [13]. The effect is related to the fact that the orbital motion of Cooper pairs yields surface currents without application of an external magnetic field, hence one can observe a Hall potential difference just because of the electric current flowing through the Hall bar.

Instead of being proportional to magnetic field H, in this case $V_{Hall} \propto N$, the chiral number. Thus, applying a DC current one can register an alternating $V_{Hall}$ because of the alternating N.

### III. ENTANGLEMENT

We consider now possible designs for quantum computing using the ruthenate qubits (RQ). The analogy between RQ and atomic qubits at $\varepsilon \neq 0$ (Fig.1) is not merely formal; they are similar enough that one can apply some known tricks borrowed from atomic QC systems including coupling with the resonator modes, selective excitation, etc. Other methodologies are emerging from contemporary superconductivity developments.

Our task is to demonstrate a conceptual approach without obvious impediments for each of QC requirements. Let us start with a qubit prepared as a thin (thickness d <1000Å) film of the ruthenate deposited on a dielectric substrate. If one such qubit is placed near another and the two are connected with a normal-metal strip, they will start interacting because of the proximity effect. The wave functions of superconductors will interfere. The usual Josephson effect is a manifestation of such interference. A non-zero Josephson current will flow between these two qubits until phases are identical, so that the system is entangled. One of the requirements for QC designs is to be able to switch this inter-qubit interaction "on" and "off" upon demand. For this purpose one can utilize the field-effect in a manner described by Fig.2.

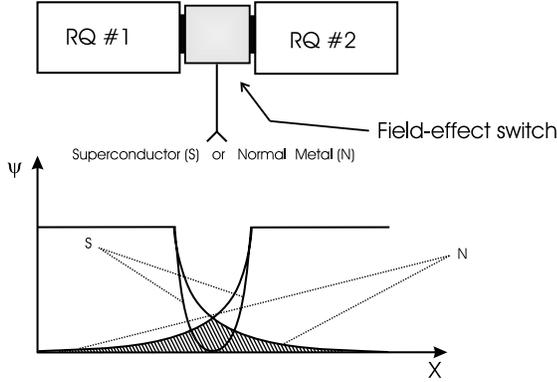

Fig. 2. Entanglement on demand. Two RQs are connected via a weak link. The overlapping of wavefunctions in case of normal-metal weak-link (dashed) disappears (s- and p-wave superconductors interfere destructively [4]) when it's turned into an ordinary superconductor via field-effect [19].

### IV. INITIAL STATE AND GATES

One can imagine at least two ways to prepare the "pure" initial stage in an RQ. The first employs thermodynamic relaxation into the lowest energy state and could be achieved at deep cooling by applying a magnetic field and splitting the degenerate levels. This way is traditional in QC. Another, more specific mechanism is described in Fig. 3. A single-RQ gate can be achieved just using dynamics of tunneling, described in Section II: it is just the unitary transformation (rotation) of the qubit state. Another approach is related to a selective action of an RF field on a single RQ. A magnetic field should be applied with a gradient along the RQ-chain, so that the resonance between the RF field and energy splitting between states |-1> and |1> takes place only for a targeted RQ. An important point is that action of the external RF field should not destroy the superconducting state, i.e., the Bose-condensate should not be 'evaporated': the applied energy amount should be less than the threshold of creation of single-particle excitations, i.e., $2\Delta$. This leads to restrictions on the volume of the RQ. For an estimate, we put $2\Delta \approx 1$meV. We also assume that the 'unitary transformation' should be performed between the states split by the magnetic field H~1G. For a single Cooper pair this is $\varepsilon = \mu H = (e\hbar/2m)H$, and for ruthenates, $m \sim 4m_e$, so that $\varepsilon = \mu H \sim 10^{-9}$ eV. This means that the allowed number of pairs $n_S = \Delta/\varepsilon$ in the Bose-ensemble is about $n_S \sim 10^6$. A unit cell of the ruthenate has a volume ~100Å$^3$, which means that the allowed volume of the qubit is $V \sim 10^8$Å$^3$, which results in typical sizes: 100×100×10 nm$^3$. These sizes are quite appropriate since they are still smaller than the magnetic penetration depth ($\lambda_L$ ~2000Å [21]), and accessible via electron lithography.

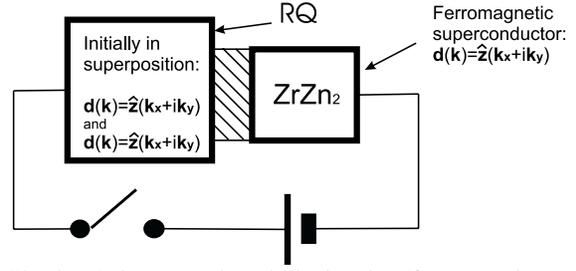

Fig. 3. Pair current through the junction 'ferromagnetic superconductor' (i.e., ZrZn$_2$ [20]) and RQ can 'prepare' the required initial state in RQ.

The controlled-Not (or C-Not) operation is related to the conditional dynamics of two qubits. One possibility is to use the analogy between the RQ and two-level 'atomic' qubit. The exchange interaction due to the proximity effect via the switched on weak-link should influence the 'energy terms' of RQs, so one might proceed with techniques already developed for 'atomic' qubits. Another opportunity is related to the fact that the C-Not gate has been shown [22] to be equivalent to a single gate plus a 'SWAP' operation (see Fig.5).

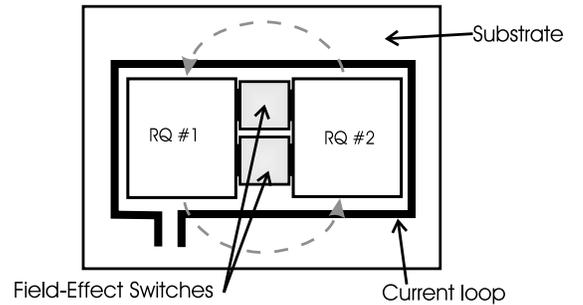

Fig. 5. Design of the SWAP-gate. For swapping one should switch the weak links 'on' and put a pulse of current into the current loop. As soon as the condensates are swapped, the weak links are 'off'.

## V. Readout and Scaleability

A way to determine the chirality of the RQ is to measure the sign of the above-mentioned spontaneous Hall-voltage [13]. Another possibility is to measure the boundary currents in the qubit. Yet one more possibility is to use the same junction with a ferromagnetic superconductor which was suggested for preparation of the initial state – it could be the simplest way to achieve readout.

Networking for chains of ruthenate qubits can be considered along the same lines as in the well-known scheme by Kane [23]. The proximity effect via switchable weak links can serve for bridging between different RQs to achieve as long a chain as is required. In addition, it is not precluded to implement a "chess-board" architecture for a scaleable design.

## VI. Additional Resources and Problems

We have shown that ruthenates are viable QC candidates. The ruthenates' critical temperature ($T_c \sim 1.5$K) is regarded as relatively low in the world of superconductivity, but in this context it should be considered as very high compared to nanokelvin temperature scale of atomic Bose-condensates. Conceptually, the physics of ruthenate qubits comes close to an area of spintronics very actively developed during the recent years [24]. Many of the effects addressed during this development can find an application in construction of RQ. For example, field-effect switching of ferromagnetism [25] can be used for suppression of superconductivity in weak links and help with interaction 'on demand' between qubits. There are other physical resources, which have not yet been employed in RQ designing. One of most interesting is the effect of disappearance of coherent effects between s- and p-wave pairing superconductors [26]. This effect is valid in absence of spin-orbital coupling and for ruthenates this coupling seems not to be strong. Preliminary investigation [27] shows some selectivity in tunneling between ordinary and triplet superconductors. Unfortunately, no one has yet achieved superconducting thin films of ruthenates. Films of triplet superconductors should be highly perfect in crystal structure to be superconductive. One purpose of this article is to motivate efforts in that direction, by showing the substantial potential benefits. Another purpose is to motivate more thorough theoretical studies of huge potential of triplet superconductivity for practical implementation of QC.

### Acknowledgment

The authors are grateful to H. Gursky, D. Van Vechten, P. Reynolds, Y. Liu, I Mazin, D. Singh and M. Lovellette for interest to this work and useful discussions. This work was supported in part by the ONR Grant N0001401WX21228.